\documentclass[apj]{emulateapj}
\usepackage{mathptmx}

\def\gtorder{\mathrel{\raise.3ex\hbox{$>$}\mkern-14mu
             \lower0.6ex\hbox{$\sim$}}}
\def\ltorder{\mathrel{\raise.3ex\hbox{$<$}\mkern-14mu
             \lower0.6ex\hbox{$\sim$}}}

\slugcomment{Draft of \today}

\shorttitle{PTF photometric catalog}

\shortauthors{Ofek et al.}

\begin{document}

\title{The Palomar Transient Factory photometric catalog 1.0}
\author{
E.~O.~Ofek\altaffilmark{1},
R.~Laher\altaffilmark{2},
J.~Surace\altaffilmark{2},
D.~Levitan\altaffilmark{3},
B.~Sesar\altaffilmark{3},
A.~Horesh\altaffilmark{3},
N.~Law\altaffilmark{4},
J.~C.~van~Eyken\altaffilmark{5},
S.~R.~Kulkarni\altaffilmark{3},
T.~A.~Prince\altaffilmark{3},
P.~Nugent\altaffilmark{6},
M.~Sullivan\altaffilmark{7},
O.~Yaron\altaffilmark{1},
A.~Pickles\altaffilmark{8},
M.~Ag\"{u}eros\altaffilmark{9},
I.~Arcavi\altaffilmark{1},
L.~Bildsten\altaffilmark{10}$^{,}$\altaffilmark{11},
J.~Bloom\altaffilmark{12},
S.~B.~Cenko\altaffilmark{12},
A.~Gal-Yam\altaffilmark{1},
C.~Grillmair\altaffilmark{2},
G.~Helou\altaffilmark{2},
M.~M.~Kasliwal\altaffilmark{1},
D.~Poznanski\altaffilmark{13},
R.~Quimby\altaffilmark{14}
}
\altaffiltext{1}{Benoziyo Center for Astrophysics, Weizmann Institute
  of Science, 76100 Rehovot, Israel.}
\altaffiltext{2}{Spitzer Science Center, MS 314-6, California Institute of Technology, Pasadena, CA 91125, USA}
\altaffiltext{3}{Division of Physics, Mathematics and Astronomy, California Institute of Technology, Pasadena, CA 91125, USA}
\altaffiltext{4}{Dunlap Institute for Astronomy and Astrophysics,
  University of Toronto, 50 St. George Street, Toronto, Ontario M5S
  3H4, Canada.}
\altaffiltext{5}{NASA Exoplanet Science Institute, California Institute of Technology, 770 South Wilson Avenue, M/S 100-22, Pasadena, CA, 91125, USA}
\altaffiltext{6}{Department of Astronomy, University of California,
  Berkeley, Berkeley, CA 94720-3411.}
\altaffiltext{7}{Department of Physics, University of Oxford, Denys
  Wilkinson Building, Keble Road, Oxford OX1 3RH, UK.}
\altaffiltext{8}{Las Cumbres Observatory Global Telescope Network, Santa Barbara, CA 93117}
\altaffiltext{9}{Columbia University, Department of Astronomy, 550 West 120th street, New York, NY 10027}
\altaffiltext{10}{Department of Physics, Broida Hall, University of
  California, Santa Barbara, CA 93106.}
\altaffiltext{11}{Kavli Institute for Theoretical Physics, Kohn Hall,
  University of California, Santa Barbara, CA 93106.}
\altaffiltext{12}{Department of Astronomy, University of California,
  Berkeley, Berkeley, CA 94720-3411.}
\altaffiltext{13}{School of Physics and Astronomy, Tel-Aviv University, Israel}
\altaffiltext{14}{IPMU, University of Tokyo, 5-1-5 Kashiwanoha, Kashiwa-shi, Chiba, 277-8583, Japan}

\begin{abstract}

We construct a photometrically calibrated catalog of
non-variable sources from the Palomar Transient Factory (PTF) observations.
The first version of this catalog presented here,
the PTF photometric catalog 1.0,
contains calibrated $R_{{\rm PTF}}$-filter magnitudes
for $\approx2.1\times10^{7}$ sources
brighter than magnitude 19,
over an area of $\approx11233$\,deg$^{2}$.
The magnitudes are provided in the PTF photometric system,
and the color of a source
is required in order to convert these magnitudes into other
magnitude systems.
We estimate that the magnitudes in this catalog
have typical accuracy of about 0.02\,mag with respect
to magnitudes from the Sloan Digital Sky Survey.
The median repeatability of
our catalog's magnitudes for stars between 15 and 16 mag,
is about 0.01\,mag, 
and it is better than 0.03\,mag for 95\% of the sources
in this magnitude range. 
The main goal of this catalog is to provide
reference magnitudes for photometric calibration
of visible light observations.
Subsequent versions of this catalog,
which will be published incrementally online, will be extended to
a larger sky area and will also include $g_{{\rm PTF}}$-filter magnitudes,
as well as variability and proper motion information.

\end{abstract}


\section{Introduction}
\label{Introduction}

All-sky photometrically-calibrated stellar catalogs
are being used to measure the true apparent flux of astrophysical sources.
Other approaches, like observing standard stars (e.g., Landolt 1992),
are time consuming since they require additional observations
(which are not of the source of interest)
under photometric conditions.
Therefore, it is desirable to have an all-sky catalog that contains
calibrated stellar magnitudes.
To date, the most widely used catalog for this purpose
is probably the USNO-B1.0 (Monet et al. 2003),
which provides the blue, red and near infra-red photographic plate magnitudes
for about $10^{9}$ sources.
Unfortunately, the photometric measurements in the USNO-B1 catalog
show significant systematic variations in the magnitude zeropoint
as a function of the position on the sky ($\sim0.5$\,mag),
even at small angular scales (Sesar et al. 2006).

The Sloan Digital Sky Survey (SDSS)
is calibrated to an accuracy of better than 2\%
(Adelman-McCarthy et al. 2008; Padmanabhan et al. 2008).
However, SDSS Data Release 8 covers only about a third of the celestial sphere.
Another possibility is to use bright Tycho-2 (H{\o}g et al. 2000)
stars to photometrically calibrate images
(Ofek 2008; Pickles \& Depagne 2010).
However, this approach requires that the Tycho stars,
brighter than magnitude $\approx12$,
are not saturated in the images.

The Palomar Transient
Factory\footnote{http://www.astro.caltech.edu/ptf/}
(PTF; Law et al. 2009; Rau et al. 2009)
is a synoptic survey designed to explore the transient sky
and to study stellar variability.
The project utilizes the $48''$ Samuel Oschin Schmidt Telescope
at Palomar Observatory.
The telescope has a digital camera equipped with 11 active
CCDs\footnote{The camera has 12 CCDs of which one is not functional.},
each 2K$\times$4K pixels (Rahmer et al. 2008),
and has been surveying the northern sky since March 2009.
Each PTF image covers 7.26\,deg$^{2}$
with a pixel scale of $1.01''$\,pix$^{-1}$.
The median point-spread function full-width at half maximum
is $\approx2''$ and it is uniform over the camera field of view
(Law et al. 2010).
The PTF main survey is currently performed in the $g$ band
during dark time and in the Mould $R$ band during bright time,
but most of the data taken prior to January 2011 were
obtained using the $R$-band filter.
In addition, a few nights around times of full Moon
are used for
surveying the sky with narrow-band H$\alpha$ filters.
An overview of the PTF survey and its first-year performance
is given in Law et al. (2010).

The PTF data are reduced by pipelines running at
Caltech's Infrared Processing and Analysis Center (IPAC).
The processing includes astrometric
and photometric calibration.
Here we build on the PTF photometric calibration to construct
a catalog of calibrated non-variable sources.
This catalog can be used to photometrically calibrate
other visible-light observations.

The paper is organized as follows.
In \S\ref{Calib}, we briefly discuss the PTF photometric calibration.
The construction of the photometric catalog is described in \S\ref{Const}.
The catalog is presented in \S\ref{Cat}
and we discuss its accuracy and repeatability in \S\ref{Acc}.
Finally we conclude in \S\ref{Conc}.

\section{Brief description of the PTF photometric calibration}
\label{Calib}

Here we briefly describe the
photometric calibration of the PTF images,
which is fundamental to the construction of the PTF photometric catalog.
A full description can be found in Ofek et al. (2012).

We use images reduced by the IPAC-PTF
pipeline (Grillmair et al. 2010; Laher et al., in prep.).
The processing includes splitting
the multi-extension FITS images,
de-biasing, flat-fielding,
astrometric calibration,
generation of mask images,
source extraction, and photometric calibration.
The astrometric calibration is performed relative to SDSS
when possible and the UCAC-3 catalog
(Zacharias et al. 2010) when SDSS information
is not available.
If a UCAC-3 solution
is not found, then the astrometry is solved
against USNO-B1.0 (Monet et al. 2003).
The median astrometric rms in single axis is
$0.11''$, $0.13''$ and $0.4''$,
for the SDSS, UCAC-3 and USNO-B1.0 catalogue, respectively.
The masks flag pixels with image artifacts,
including ghosts, halos, aircraft/satellite tracks,
saturation, CCD bleeding, and dead/bad pixels.
Sources which contain masked pixels inherit
the pixels' flag,
and these are stored in the catalogs
associated with the processed images.

Our photometric calibration method
is similar to the classical method of observing
standard stars through various airmasses
and assuming photometric conditions --
i.e., the atmospheric transmission properties are constant in time
and are a continuous function of airmass.
On average, we                   
typically observe $\sim10^{5}$ SDSS stars with high signal-to-noise ratio (S/N)
per CCD per night.
Therefore, we usually have a sufficient number of photometric
measurements to robustly constrain all calibration parameters
for a given night.

After selecting high S/N sources, we fit
the difference between the instrumental magnitude measured in
the PTF system (e.g., $R_{{\rm PTF}}^{{\rm inst}}$),
and the standard star magnitudes measured by SDSS
(e.g., $r_{{\rm SDSS}}$) with a simple model.
The free parameters in this
model include the global zero point of the image,
color term, extinction coefficient, color-airmass term,
variation of the global zero point during the night,
exposure time,
and the illumination correction.
Here the illumination correction represents
variation of the photometric zero point as a function of position
on the CCD.
The latter correction is represented by two
methods relative to the center of each CCD,
which are described in Ofek et al. (2012).
To generate the first version of the PTF
photometric catalog, we use the model
in which the 
variation of the photometric zero point with CCD
position is a two-dimensional, low-order polynomial
in the position
across each CCD (i.e., Eq.~3 in Ofek et al. 2012).
The goodness of the fit is described by several
estimators, including
the RMS of residuals of bright stars
from the best fit model
(parameter APBSRMS in Table~2 in Ofek et al. 2012).

\begin{figure}
\centerline{\includegraphics[width=8.5cm]{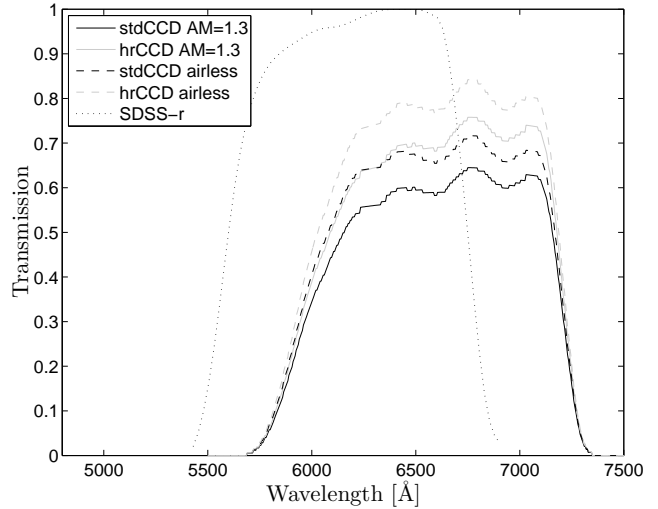}}
\caption{PTF $R$-band filter transmission at airmass of 1.3
(solid line) and no atmosphere (dashed line).
The transmission is shown for the two CCD types in the PTF camera.
'std' stands for the standard CCD
(black line; CCDID 0,1,2,6,7,8,9,10,11)
while 'hr' stands for the high resistivity CCDs
(gray line; CCDID 4,5).
The transmission was calculated by multiplying
the filter, CCD and atmospheric transmissions.
The atmospheric transmission was calculated using
a standard smooth atmosphere (Hayes \& Latham 1975) for 1.7\,km elevation
and airmass $1.3$.
For reference we also show the transmission of the SDSS $r$-band filter.
\label{fig:R_FilterTran}}
\end{figure}

We note that the magnitudes produced
by our calibration process are related to
the SDSS magnitude system (and other systems) 
via the aforementioned color terms.
In the case where the $r-i$ color of an object is known,
it is possible to convert the PTF $R$-band magnitudes
to other systems.
These transformations are described in
Ofek et al. (2012; Eqs.~4--7),
while the color terms for the 11 active CCDs
are given in Ofek et al. (2012; Table~3).
For reference we present the PTF $R$-band filter transmission
in Figure~\ref{fig:R_FilterTran} and in Table~\ref{Tab:R_FilterTran}.
Subsequent versions of the PTF photometric catalog
will additionally include $g_{{\rm PTF}}$ magnitudes,
which will enable straightforward transformation
into other magnitude systems.
\begin{deluxetable}{lllllll}
\tablecolumns{7}
\tabletypesize{\scriptsize}
\tablewidth{0pt}
\tablecaption{PTF $R$-band filter transmission}
\tablehead{
\colhead{$\lambda$} &
\colhead{Filter}     &
\colhead{QE$_{{\rm std}}$}     &
\colhead{QE$_{{\rm hr}}$}      &
\colhead{Atm.} &
\colhead{Sys$_{{\rm std}}$} &
\colhead{Sys$_{{\rm hr}}$}  \\
\colhead{\AA} &
\colhead{}    &
\colhead{}    &
\colhead{}    &
\colhead{}    &
\colhead{}    &
\colhead{}
}
\startdata
5680.0 & 0.00 & 0.66 & 0.74 & 0.84 & 0.00 & 0.00\\
5685.0 & 0.00 & 0.66 & 0.74 & 0.84 & 0.00 & 0.00\\
5690.0 & 0.01 & 0.66 & 0.74 & 0.84 & 0.01 & 0.01\\
5695.0 & 0.01 & 0.66 & 0.75 & 0.84 & 0.01 & 0.01\\
5700.0 & 0.01 & 0.67 & 0.75 & 0.84 & 0.01 & 0.01
\enddata
\tablecomments{PTF $R$-band filter transmission
(see also Figure~\ref{fig:R_FilterTran}).
$\lambda$ is the wavelength,
'Filter' is the filter transmission,
'QE' is the CCD efficiency,
'Atm.' is the atmosphere transmission calculated using
a standard smooth atmosphere (Hayes \& Latham 1975) for 1.7\,km elevation
and airmass $1.3$,
and 'Sys' is the total efficiency calculated
by multiplying the filter, QE and the atmosphere transmissions.
Subscript 'std' stands for standard CCD, while 'hr' stands for
high resistivity (see Figure~\ref{fig:R_FilterTran} caption).
This table is published in its entirety in the electronic edition of
{\it PASP}. A portion of the full table is shown here for
guidance regarding its form and content.}
\label{Tab:R_FilterTran}
\end{deluxetable}

\section{Catalog construction}
\label{Const}

Version 1.0 of the
catalog was constructed from PTF data taken
before November 2011, using the IPAC-PTF pipeline software version
identifier (SVID)$>47$.
For this version of the catalog, we only used
PTF fields that were observed with the Mould $R$ band
on at least three photometric nights under good conditions.
The terms ``photometric nights'' and
``good conditions'' are not well defined
and depend on the required photometric accuracy.
Here we select only images 
that have photometric and
quality parameters within the ranges
specified in Table~\ref{Tab:Pars}.

A detailed description of the photometric parameters
(i.e., APBSRMS, $\alpha_{{\rm c},R}$, $\alpha_{{\rm a},R}$)
and their distributions
is available in Ofek et al. (2012).
Here $MoonESB$ is the theoretical $V$-band
excess in sky surface magnitude (negative number)
due to the Moon.
This excess is calculated
using the algorithm of Krisciunas \& Schaefer (1991).
$APBSRMS$ is the root mean square (RMS) of bright stars
of the nightly photometric
calibration residuals from the best fit,
$\alpha_{{\rm c},R}$ is the $R$-band $r-i$-color term coefficient
of the nightly photometric solution,
and 
$\alpha_{{\rm a},R}$ is the nightly $R$-band extinction coefficient.
The allowed ranges of these parameters (listed in Table~\ref{Tab:Pars})
correspond to $\pm3$ times the one standard deviation\footnote{We use the 68 percentile divided by two as a robust estimator for one standard deviation.}
from the median value of each parameter over all data taken with
a given CCD\footnote{The CCD number is designated by CCDID, which ranges from 0 to 11 (CCDID=3 is inoperable).}.
\begin{deluxetable}{llrrl}
\tablecolumns{5}
\tabletypesize{\scriptsize}
\tablewidth{0pt}
\tablecaption{Ranges of parameters of good data}
\tablehead{
\colhead{Parameter} &
\colhead{CCDID} &
\colhead{Min}       &
\colhead{Max}       &
\colhead{Units}
}
\startdata
Seeing             &all  & \nodata & $4.0$   & arcsec            \\
$MoonESB$          &all  & $-3.0$  & \nodata & mag\,arcsec$^{-2}$ \\
$APBSRMS$          &all  & \nodata & $0.04$  & mag               \\
$\alpha_{{\rm c,R}}$ & 0   & 0.190   & 0.244   & mag\,mag$^{-1}$    \\
$\alpha_{{\rm a,R}}$ & 0   & $-0.182$&$-0.044$ & mag\,airmass$^{-1}$\\
$\alpha_{{\rm c,R}}$ & 1   & 0.175   & 0.235   & mag\,mag$^{-1}$  \\
$\alpha_{{\rm a,R}}$ & 1   & $-0.177$&$-0.039$ & mag\,airmass$^{-1}$ \\ 
$\alpha_{{\rm c,R}}$ & 2   & 0.178   & 0.226   & mag\,mag$^{-1}$  \\
$\alpha_{{\rm a,R}}$ & 2   & $-0.187$&$-0.037$ & mag\,airmass$^{-1}$ \\ 
$\alpha_{{\rm c,R}}$ & 4   & 0.189   & 0.249   & mag\,mag$^{-1}$  \\
$\alpha_{{\rm a,R}}$ & 4   & $-0.200$&$-0.038$ & mag\,airmass$^{-1}$ \\ 
$\alpha_{{\rm c,R}}$ & 5   & 0.200   & 0.254   & mag\,mag$^{-1}$  \\
$\alpha_{{\rm a,R}}$ & 5   & $-0.198$&$-0.036$ & mag\,airmass$^{-1}$ \\ 
$\alpha_{{\rm c,R}}$ & 6   & 0.201   & 0.249   & mag\,mag$^{-1}$  \\
$\alpha_{{\rm a,R}}$ & 6   & $-0.183$&$-0.027$ & mag\,airmass$^{-1}$ \\ 
$\alpha_{{\rm c,R}}$ & 7   & 0.172   & 0.238   & mag\,mag$^{-1}$  \\
$\alpha_{{\rm a,R}}$ & 7   & $-0.177$&$-0.033$ & mag\,airmass$^{-1}$ \\ 
$\alpha_{{\rm c,R}}$ & 8   & 0.181   & 0.229   & mag\,mag$^{-1}$  \\
$\alpha_{{\rm a,R}}$ & 8   & $-0.175$&$-0.049$ & mag\,airmass$^{-1}$ \\ 
$\alpha_{{\rm c,R}}$ & 9   & 0.165   & 0.237   & mag\,mag$^{-1}$  \\
$\alpha_{{\rm a,R}}$ & 9   & $-0.188$&$-0.038$ & mag\,airmass$^{-1}$ \\ 
$\alpha_{{\rm c,R}}$ & 10  & 0.176   & 0.260   & mag\,mag$^{-1}$  \\
$\alpha_{{\rm a,R}}$ & 10  & $-0.188$&$-0.038$ & mag\,airmass$^{-1}$ \\ 
$\alpha_{{\rm c,R}}$ & 11  & 0.188   & 0.248   & mag\,mag$^{-1}$  \\
$\alpha_{{\rm a,R}}$ & 11  & $-0.189$&$-0.039$ & mag\,airmass$^{-1}$ 
\enddata
\tablecomments{Min and Max specify the range minimum and maximum, respectively. See text for details.}
\label{Tab:Pars}
\end{deluxetable}

We choose PTF fields\footnote{A PTF field, denoted by PTFFIELD, is uniquely associated with a predefined sky position.}
that have at least three images
taken on three different nights, with the
criteria listed in Table~\ref{Tab:Pars}.
The requirement to analyze only fields that were
observed on three or more photometric nights is important
in order to remove outliers that may be present in
the data.
For example, if a night was photometric for
90\% of the time (e.g., clouds entered toward the end of the night),
then our pipeline might claim that the night was photometric, but the calibration
of some of the data will be poor.
Therefore, in order to get the calibrated source magnitudes,
it is important to average the data taken over several
photometric nights.
Moreover, observations taken on multiple nights
allow us to calculate variability and proper motion indicators.

\begin{figure}
\centerline{\includegraphics[width=8.5cm]{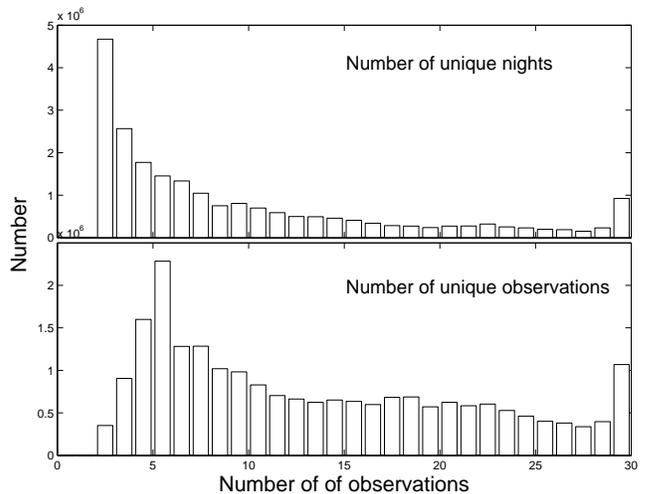}}
\caption{Histogram of the number of unique nights (upper panel)
and number of unique observations (lower panel)
per object in the catalog.
The histogram of number of unique observations peaks
at six observations, twice of the value at which the histogram
of number of unique night peaks. This reflects the fact
that until March 2012 for most fields we obtained two observations
per night.
\label{fig:HistN}}
\end{figure}

In order to expedite the processing
in cases where we have
more than 30 images of the same field,
we truncated the number of images according to
the following scheme.
If more than 30 images of the same field,
taken on less than 30 unique nights,
were available, then we selected a single random image
from each night.
If more than 30 images taken in more than 30 unique nights
were available,
then we selected only the 30 nights with the smallest
APBSRMS parameter and selected one image
from each one of these nights.
Histograms of the number of unique nights and
number of unique observations per object in the catalog are
shown in Figure~\ref{fig:HistN}.

For each set of selected images
of a given PTFFIELD/CCDID,
we matched the sources in all the images
against a reference image
with a matching radius of $1.5''$.
Here, the reference image was selected
as the image of the field with the largest number of
sources\footnote{Typically, this is the image with the best limiting magnitude.}.
We note that in future catalog versions, we intend to use
a deep coadd image for each PTFFIELD/CCDID as a reference image.

Next, we remove all the measurements that are masked
by one of the following flags:
source is deblended by SExtractor (Bertin \& Arnouts 1996);
aircraft/satellite track;
high dark current;
noisy/hot pixel;
containing possible optical ghost;
CCD-bleed;
radiation hit\footnote{The current radiation hit/CCD bleed flag in the version of the PTF IPAC pipeline used here is not a good indicator for radiation hits.};
saturated pixels;
dead pixel;
not a number;
halo around bright star;
dirt on optics.
These flags are described in details in Laher et al. (in prep).
The remaining photometric measurements are used to calculate
the mean photometric properties of each source
in the PTF photometric catalog 1.0.

Some of the PTF fields are overlapping in areal coverage.
Therefore, some of the sources generated
by the process described above are duplicates.
We remove duplicate sources and keep the catalog entry
which corresponds to the lowest PTFFIELD.
We also remove all sources fainter than 19 mag.
Sources fainter than $\sim19$\,mag have photometric errors that are
larger than the photometric accuracy of this catalog.

\section{The catalog}
\label{Cat}

The PTF photometric catalog is
presented in Table~\ref{Tab:Cat}.
The catalog is also available online
from the IPAC website\footnote{http://irsa.ipac.caltech.edu/}
and the VizieR service\footnote{http://vizier.u-strasbg.fr/viz-bin/VizieR}.
The following columns are available:
\begin{deluxetable*}{llllllllllllll}
\tablecolumns{14}
\tabletypesize{\scriptsize}
\tablewidth{0pt}
\tablecaption{PTF photometric catalog 1.0}
\tablehead{
\colhead{$\alpha_{{\rm J2000}}$} &
\colhead{$\delta_{{\rm J2000}}$} &
\colhead{$N_{{\rm obs}}$} &
\colhead{$N_{{\rm night}}$} &
\colhead{$Best RMS$} &
\colhead{$R_{{\rm PTF}}$} &
\colhead{$\Delta{R_{{\rm PTF}}}$} &
\colhead{$\Delta_{-}{R_{{\rm PTF}}}$} &
\colhead{$\Delta_{+}{R_{{\rm PTF}}}$} &
\colhead{$\mu_{{\rm type}}$} &
\colhead{$\Delta\mu_{{\rm type}}$} &
\colhead{PTFFIELD} &
\colhead{CCDID} &
\colhead{Flag} \\
\colhead{deg} &
\colhead{deg} &
\colhead{} &
\colhead{} &
\colhead{mag} &
\colhead{mag} &
\colhead{mag} &
\colhead{mag} &
\colhead{mag} &
\colhead{mag} &
\colhead{mag} &
\colhead{} &
\colhead{} &
\colhead{} 
}
\startdata
 42.532500& $-31.087161$ &  23 & 12 & 0.022 &  16.152& 0.021  &0.017& 0.026&    0.058&  0.095&   100111& 09&   1\\
 42.786188& $-31.086008$ &  21 & 12 & 0.022 &  17.172& 0.030  &0.022& 0.037&    0.058&  0.060&   100111& 09&   1\\
 42.593730& $-31.085419$ &  22 & 12 & 0.022 &  18.507& 0.063  &0.055& 0.071&    0.106&  0.106&   100111& 09&   1\\
 42.525416& $-31.081556$ &  23 & 12 & 0.022 &  14.225& 0.015  &0.017& 0.014&    0.046&  0.072&   100111& 09&   1\\
 42.953363& $-31.080395$ &  23 & 12 & 0.022 &  15.286& 0.025  &0.029& 0.022&    0.003&  0.063&   100111& 09&   1
\enddata
\tablecomments{The table is sorted by declination. See text for column descriptions. This table is published in its entirety in the electronic edition of
{\it PASP}. A portion of the full table is shown here for
guidance regarding its form and content.}
\label{Tab:Cat}
\end{deluxetable*}

\bigskip

\noindent
$\alpha_{{\rm J2000}}$: The median J2000.0 right ascension
over all the images used to derive the photometry.

\noindent
$\delta_{{\rm J2000}}$: The median J2000.0 declination
over all the images used to derive the photometry.

\noindent
$N_{{\rm obs}}$: Number of images that were used
to derive the photometry.

\noindent
$N_{{\rm night}}$: Number of individual nights in
which the images were taken.

\noindent
$BestRMS$: The best bright-star RMS value (APBSRMS parameter)
over all nights used.

\noindent
$R_{{\rm PTF}}$: The median $R$-band magnitude in the PTF system
(i.e., not color corrected to SDSS) over all epochs.

\noindent
$\Delta{R_{{\rm PTF}}}$: The error on the median $R$-band
magnitude as calculated from
the range containing 68\% of the measurements divided by two
(i.e., $[\Delta_{-}{R_{{\rm PTF}}} + \Delta_{+}{R_{{\rm PTF}}}]/2$).

\noindent
$\Delta_{-}{R_{{\rm PTF}}}$: The lower-bound error on the median $R$-band
magnitude as calculated from
the range between the median magnitude and
the lower 16-percentile magnitude.

\noindent
$\Delta_{+}{R_{{\rm PTF}}}$: The upper-bound error on the median $R$-band
magnitude as calculated from
the range between the median magnitude and
the upper 16-percentile magnitude.
By construction the sum of $\Delta_{-}{R_{{\rm PTF}}}$ and
$\Delta_{+}{R_{{\rm PTF}}}$ gives the 68-percentile range,
and it is therefore an estimator for twice the 1-$\sigma$ error.

\noindent
$\mu_{{\rm type}}$: An indicator that can be used to estimate if the source
is resolved (e.g., galaxy) or unresolved (e.g., star).
This is based on the SExtractor parameters MAG\_AUTO$-$MU\_MAX. MU\_MAX measures
the object surface magnitude in the object's central pixel.
The parameter is normalized such that the median
of $\mu_{{\rm type}}$ over the population of sources
in the image is zero.
Since most sources in PTF images are unresolved (stellar-like),
this parameter has a value near zero for unresolved sources,
and extended sources will have negative $\mu_{{\rm type}}$ values.
Based on our preliminary calibration, there is a
$\sim2\%$ probability that 
objects with $\mu_{{\rm type}}<-0.2$ are unresolved (stellar).

\noindent
$\Delta\mu_{{\rm type}}$: The standard deviation
of the $\mu_{{\rm type}}$ as measured over
all images used in the processing.

\noindent
{\it PTFFIELD}: The PTF unique field identifier.

\noindent
{\it CCDID}: The PTF CCD identifier (0 to 11).
 
\noindent
{\it Flag}: A flag that is set to 1 if the 68-percentile range divided by two
(i.e., $[\Delta_{-}{R_{{\rm PTF}}}+\Delta_{+}{R_{{\rm PTF}}}]/2$)
is smaller than $0.03\times10^{-0.2(14-R_{{\rm PTF}})}$
and is set to 0 otherwise.
This flag can be useful in selecting sources for which
the photometry is reliable and are likely not
strongly variable sources.
In the current version of the catalog we list only sources with $Flag=1$.

\begin{figure}
\centerline{\includegraphics[width=8.5cm]{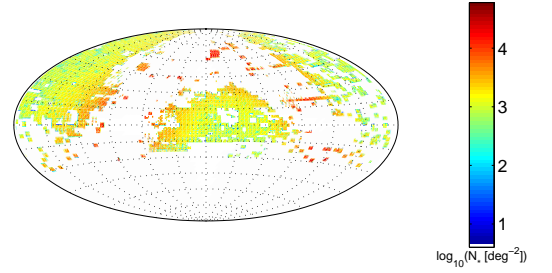}}
\caption{Coverage of the PTF photometric catalog 1.0
shown in an equal-area Aitoff projection in equatorial coordinates.
$RA=0$\,deg, $Dec=0$\,deg is in the center of the map.
The color shading shows the density of stars per deg$^{2}$,
as calculated in a grid of $0.5\times0.5$\,deg$^{2}$ cells on the sky.
\label{fig:PTFPC1_SkyDen}}
\end{figure}

The current catalog covers 11233\,deg$^{2}$ and its
sky coverage is shown in Figure~\ref{fig:PTFPC1_SkyDen}.
About 7978\,deg$^{2}$ are found within the footprint of
SDSS-DR8, while 3255\,deg$^{2}$ are outside the footprint
of SDSS-DR8.

\section{Accuracy and repeatability}
\label{Acc}

\begin{figure}
\centerline{\includegraphics[width=8.5cm]{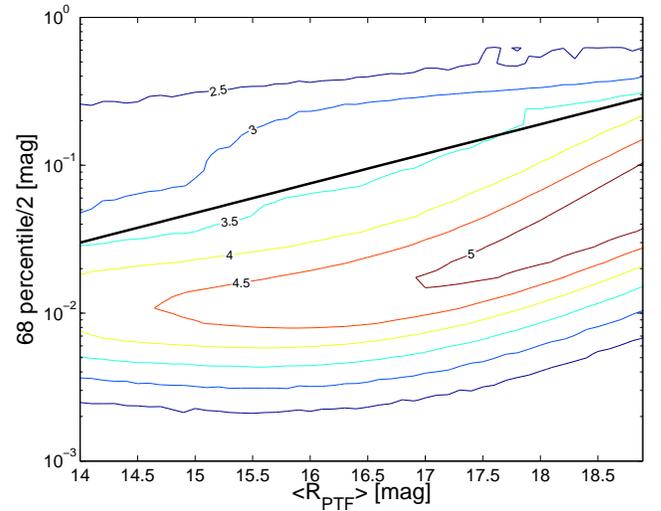}}
\caption{The 68 percentile range of the calibrated magnitude measurements
divided by two as a function of the median magnitude of each star
in the catalog.
The contours indicate the $log_{10}$ of the density
of stars in this plane as calculated in cells of 0.1\,mag in the X-axis
and 0.1\,dex in the Y-axis.
The thick-black line shows the error threshold we used
to select photometric calibrators (i.e., {\it Flag} in Table~\ref{Tab:Cat}).
The figure demonstrates that the typical calibration error for bright stars
is about 1\%--2\% and that the errors increase to about 0.06\,mag for
magnitude $\sim19$. We note that while the 1\% error at the bright
end is mostly systematic, the $\approx 6\%$ error at the faint end
is mostly statistical (Poisson errors).
2.6\% of the stars are found above the solid line
(i.e., stars with $Flag=0$ which are not listed in the current
version of the catalog). It is likely that a large fraction
of the stars above the solid line are variable stars.
\label{fig:MagSigma_Comp}}
\end{figure}

\begin{figure}
\centerline{\includegraphics[width=8.5cm]{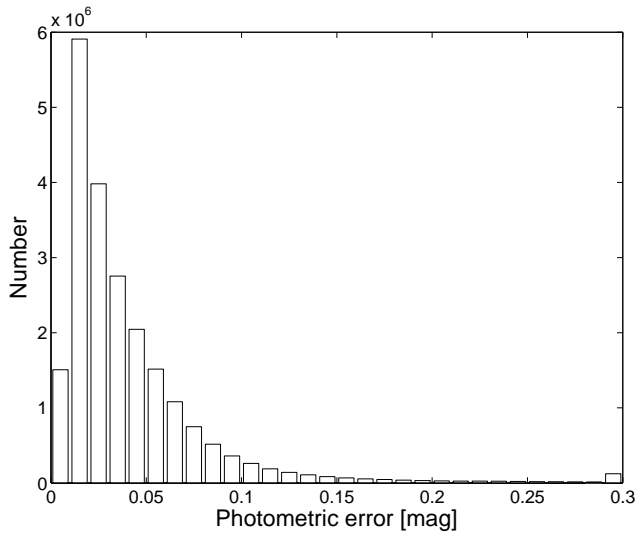}}
\caption{Histogram of the 68 percentile range divided by two (robust errors)
distribution of all the stars in the PTF photometric catalog.
The bins show the number of stars per 0.01 mag bins.
\label{fig:HistErr}}
\end{figure}

Figure~\ref{fig:MagSigma_Comp} gives contours of
objects density in the magnitude-scatter plane,
where scatter is the 68 percentile range
of the calibrated magnitude measurements divided by two.
The thick solid black line shows the error threshold we used
to select photometric calibrators (i.e., $Flag=1$ in Table~\ref{Tab:Cat}).
Figure~\ref{fig:HistErr} presents the distribution
of the 68 percentile range divided by two
of all the stars in the PTF photometric catalog.
This histogram suggest that 23.6\% (4.3\%) of the objects
in the catalog have errors worse than 5\% (10\%).
We note that fainter objects have larger photometric
(Poisson) errors. Therefore these errors do~not
represent the floor of systematic errors that can
be achieved by using bright stars or many faint stars.
\begin{figure}
\centerline{\includegraphics[width=8.5cm]{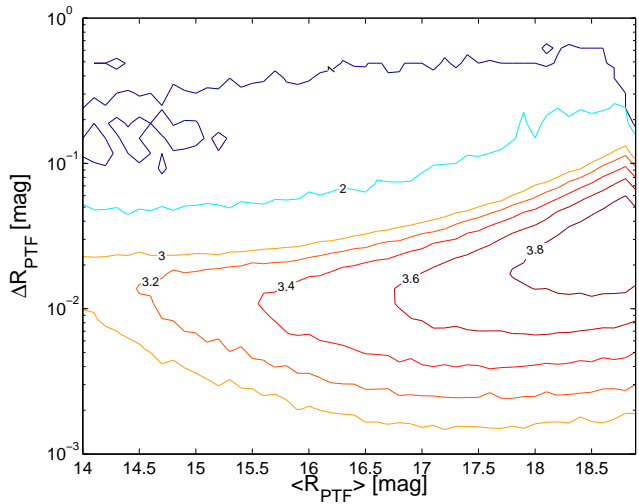}}
\caption{The differences between all
the duplicate measurements of the same stars
taken on different fields or CCDs,
as a function of their mean magnitude.
This plot is based on about $2.1\times10^{6}$ duplicate
measurements.
The contours indicate the $log_{10}$ of the density
of stars in this plane as calculated in cells of 0.1\,mag in the X-axis
and 0.1\,dex in the Y-axis.
The figure demonstrates that the typical repeatability of
the photometric catalog is about 1\%--2\% at the bright end
and 3\% at the faint end, and that in some cases repeatability
of a few mmag is achieved even without applying relative
photometry techniques.
See additional discussion in Fig.~\ref{fig:MagSigma_Comp}.
\label{fig:MultCat_Comp}}
\end{figure}

As discussed in \S\ref{Cat}, some of the sources have duplicate
measurements on different PTF fields and/or CCDs.
Such duplicate measurements can be used to test the repeatability
of the catalog over multiple CCDs.
Figure~\ref{fig:MultCat_Comp} shows the differences between all
$2.1\times10^{6}$ duplicate pairs as a function of their mean magnitude.
We note that the mean difference between two data points
generated from a Gaussian distribution is equal to 
1.13 times the standard deviation of the Gaussian.
This figure suggests that our photometric calibration
does~not depend on which CCD in the camera the data
were obtained.
In the magnitude range 15 to 16, the median repeatability
is about 0.01\,mag and 95\% of the sources have a repeatability
better than about 0.03\,mag.

\section{Conclusions}
\label{Conc}

To summarize,
we present a catalog of calibrated PTF $R$-band magnitudes
of sources extracted from PTF images.
The catalog covers about 28\% of the celestial
sphere, some of it outside the SDSS footprint.
Conversion of PTF $R$-band magnitude
to other magnitude systems requires knowledge
of the source's color.
We note that 
the scatter in colors of some populations of objects
is small enough (e.g., RR~Lyr stars, asteroids)
that their mean color can be used for conversion.
We note that the current version of the catalog is designed as a
photometric catalog, rather than astrometric catalog.

Future versions of this catalog
will also provide the $g$-band magnitudes,
which will allow one to apply color corrections directly,
with no assumptions.
In future versions,
we also plan to include more robust variability information,
source morphology and proper motion measurements of individual
sources.

\acknowledgments

We thank an anonymous referee for constructive comments.
This paper is based on observations obtained with the
Samuel Oschin Telescope as part of the Palomar Transient Factory
project, a scientific collaboration between the
California Institute of Technology,
Columbia University,
Las Cumbres Observatory,
the Lawrence Berkeley National Laboratory,
the National Energy Research Scientific Computing Center,
the University of Oxford, and the Weizmann Institute of Science.
EOO is incumbent of
the Arye Dissentshik career development chair and
is greatful to support by
a grant from the Israeli Ministry of Science.
SRK and his group are partially supported by the
NSF grant AST-0507734.


\begin{thebibliography}{}

\bibitem[Adelman-McCarthy et al.(2008)]{2008ApJS..175..297A} 
Adelman-McCarthy, J.~K., et al.\ 2008, ApJS, 175, 297 



\bibitem[Bertin \& Arnouts(1996)]{1996A&AS..117..393B} Bertin, E., \& Arnouts, S.\ 1996, A\&AS, 117, 393 







\bibitem[Grillmair et al.(2010)]{2010ASPC..434...28G} Grillmair, C.~J., et 
al.\ 2010, Astronomical Data Analysis Software and Systems XIX, 434, 28 

\bibitem[Hayes 
\& Latham(1975)]{1975ApJ...197..593H} Hayes, D.~S., \& Latham, D.~W.\ 1975, \apj, 197, 593 

\bibitem[H{\o}g et al.(2000)]{2000A&A...355L..27H} H{\o}g, E., et al.\ 
2000, A\&A, 355, L27 



\bibitem[Krisciunas 
\& Schaefer(1991)]{1991PASP..103.1033K} Krisciunas, K., \& Schaefer, B.~E.\ 1991, PASP, 103, 1033 

\bibitem[Landolt(1992)]{1992AJ....104..340L} Landolt, A.~U.\ 1992, AJ, 
104, 340 

\bibitem[Law et al.(2009)]{2009PASP..121.1395L} Law, N.~M., et al.\ 2009, 
PASP, 121, 1395 

\bibitem[Law et al.(2010)]{2010SPIE.7735E.122L} Law, N.~M., et al.\ 2010, SPIE, 7735



\bibitem[Monet et al.(2003)]{2003AJ....125..984M} Monet, D.~G., et al.\ 
2003, AJ, 125, 984 

\bibitem[Ofek(2008)]{2008PASP..120.1128O} Ofek, E.~O.\ 2008, PASP, 120, 
1128 


\bibitem[Ofek et al.(2012)]{2011arXiv1112.4851O} Ofek, E.~O., Laher, R., 
Law, N., et al.\ 2012, arXiv:1112.4851 

\bibitem[Padmanabhan et al.(2008)]{2008ApJ...674.1217P} Padmanabhan, N., et 
al.\ 2008, ApJ, 674, 1217 


\bibitem[Pickles \& Depagne(2010)]{2010PASP..122.1437P} Pickles, A., \& Depagne, {\'E}.\ 2010, PASP, 122, 1437 


\bibitem[Rahmer et al.(2008)]{2008SPIE.7014E.163R} Rahmer, G., Smith, R., 
Velur, V., Hale, D., Law, N., Bui, K., Petrie, H., 
\& Dekany, R.\ 2008, SPIE, 7014

\bibitem[Rau et al.(2009)]{2009PASP..121.1334R} Rau, A., et al.\ 2009, 
PASP, 121, 1334 

\bibitem[Sesar et al.(2006)]{2006AJ....131.2801S} Sesar, B., Svilkovi{\'c}, 
D., Ivezi{\'c}, {\v Z}., et al.\ 2006, AJ, 131, 2801 




\bibitem[York et al.(2000)]{2000AJ....120.1579Y} York, D.~G., et al.\ 2000, AJ, 120, 1579 



\bibitem[Zacharias et al.(2010)]{2010AJ....139.2184Z} Zacharias, N., Finch, 
C., Girard, T., et al.\ 2010, AJ, 139, 2184 

\end{thebibliography}
\end{document}